\documentclass[aps,floats,twocolumn,showpacs]{revtex4-1}
\usepackage{amssymb}
\usepackage{amsmath}
\usepackage{epstopdf}
\usepackage{graphicx}

 \def\bc{\begin{center}}          \def\ec{\end{center}}
 \newcommand{\I}[1]{}
 \allowdisplaybreaks
\begin{document}
 \title{Parameter sensitivity of plasma wakefields driven by self-modulating proton beams}
 \author{K.V.Lotov, V.A.Minakov, A.P.Sosedkin}
 \affiliation{Novosibirsk State University, 630090, Novosibirsk, Russia}
 \affiliation{Budker Institute of Nuclear Physics SB RAS, 630090, Novosibirsk, Russia}
 \date{\today}
 \begin{abstract}
The dependence of wakefield amplitude and phase on beam and plasma parameters is studied in the parameter area of interest for self-modulating proton beam-driven plasma wakefield acceleration. The wakefield sensitivity to small parameter variations reveals the expected level of shot-to-shot jitter of experimental results. Of all the parameters, the plasma density stands out, as the wakefield phase is extremely sensitive to this parameter. The study of large variations determines the effects that limit the achievable accelerating field in different parts of the parameter space: nonlinear elongation of the wakefield period, insufficient charge of the drive beam, emittance-driven beam divergence, and motion of plasma ions.
 \end{abstract}
 \maketitle

\section{Introduction}

The concept of proton driven plasma wakefield acceleration emerged five years ago \cite{NatPhys9-363,PRST-AB13-041301}. The main motivation for using protons is that the energy content of state-of-the-art proton beams is sufficiently high to bring a substantial amount of electrons to TeV energy scale in a single plasma section. Since plasmas can support extremely strong accelerating fields \cite{RMP81-1229}, this concept may open a path to the future of high-energy particle physics \cite{AWAKE}.

In the initial proposal \cite{NatPhys9-363,PRST-AB13-041301}, the proton beam was assumed to be longitudinally compressed to the sub-millimeter length. That was necessary for driving the plasma wave with the wavelength of 1\,mm or shorter. However, as available proton beams have lengths of about 10\,cm, such strong compression seems unrealistic to obtain with conventional techniques \cite{IPAC10-4395}. Soon it was realized that the plasma wave can slice the initially long proton beam into the train of equally spaced micro-bunches which in turn resonantly drives the plasma wave \cite{PPCF53-014003,PRL104-255003}. The slicing is caused by the saturated self-modulation instability \cite{EPAC98-806,PRL104-255003,PRL107-145003,PRL107-145002}.

\begin{table}[b]
 \caption{ Parameters of the baseline AWAKE scenario and the notation.}\label{t1}
 \bc\begin{tabular}{ll}\hline
  Parameter, notation & Value \\ \hline
  Plasma density, $n_0$ & $7 \times 10^{14}\,\text{cm}^{-3}$ \\
  Atomic weight of plasma ions, $M_i$ & 85.5 \\
  Beam population, $N_b$ & $3\times 10^{11}$ \\
  Beam length, $\sigma_{zb}$ & 12\,cm \\
  Beam radius, $\sigma_{rb}$ & 0.02\,cm \\
  Beam energy, $W_b$ & 400\,GeV \\
  Beam energy spread, $\delta W_b$ & 0.35\% \\
  Beam angular spread, $\delta \alpha_b = \epsilon_b/\sigma_{rb}$, & $4.5 \times 10^{-5}$ \\
  Seed location relative to beam center, $\xi_s$ \quad & 0\,cm \\
  \hline
  Plasma skin depth, $c/\omega_p$, & 0.02\,cm \\
  Wavebreaking field, $E_0=mc\omega_p/e$, & 2.54\,GV/m \\
  Interaction length, $L_\text{max}$ & 10\,m \\
  Maximum beam density, $n_{b0}$ & $4\times 10^{12}\,\text{cm}^{-3}$ \\
  Beam emittance, $\epsilon_b$ & $9\,\mu$m\,mrad \\
  Beam normalized emittance, $\epsilon_{bn}$ & 3.6\,mm\,mrad \\ \hline
 \end{tabular}\ec
\end{table}

To test beam self-modulation and subsequent wakefield excitation, a proof-of-principle experiment named AWAKE was launched at CERN \cite{AWAKE,IPAC13-1179,TDR,NIMA-740-48}. In the baseline experiment scenario, the 400\,GeV proton beam from the SPS synchrotron passes through the 10 meter long plasma section. Beam self-modulation is seeded by a short laser pulse co-propagating with the beam, which instantly ionizes a highly uniform rubidium vapor and produces the plasma of the same density as that of the neutral vapor \cite{Oz}. As the wakefield of the seed perturbation is substantially stronger than the shot noise \cite{PRST-AB16-041301}, the beam self-modulates at first 4~meters and then excites the plasma wave and accelerates externally injected test electrons to the energy of about 2\,GeV. The drive beam and plasma parameters corresponding to the baseline scenario are given in Table~\ref{t1}.

In this paper we numerically study how sensitive are the excited wakefields to variations of the parameters listed in the upper part of Table~\ref{t1}. The lower part values are provided for reference. We examine the wakefield amplitude in Section~\ref{s2} and the wakefield phase in Section~\ref{s3}. The wakefield response to small parameter variations characterizes the level of shot-to-shot jitter of experimental results. The study of large variations determines which of the effects can limit the accelerating fields and which parameters are to be improved for better performance.

\begin{figure*}[thb]
\bc\includegraphics[width=453bp]{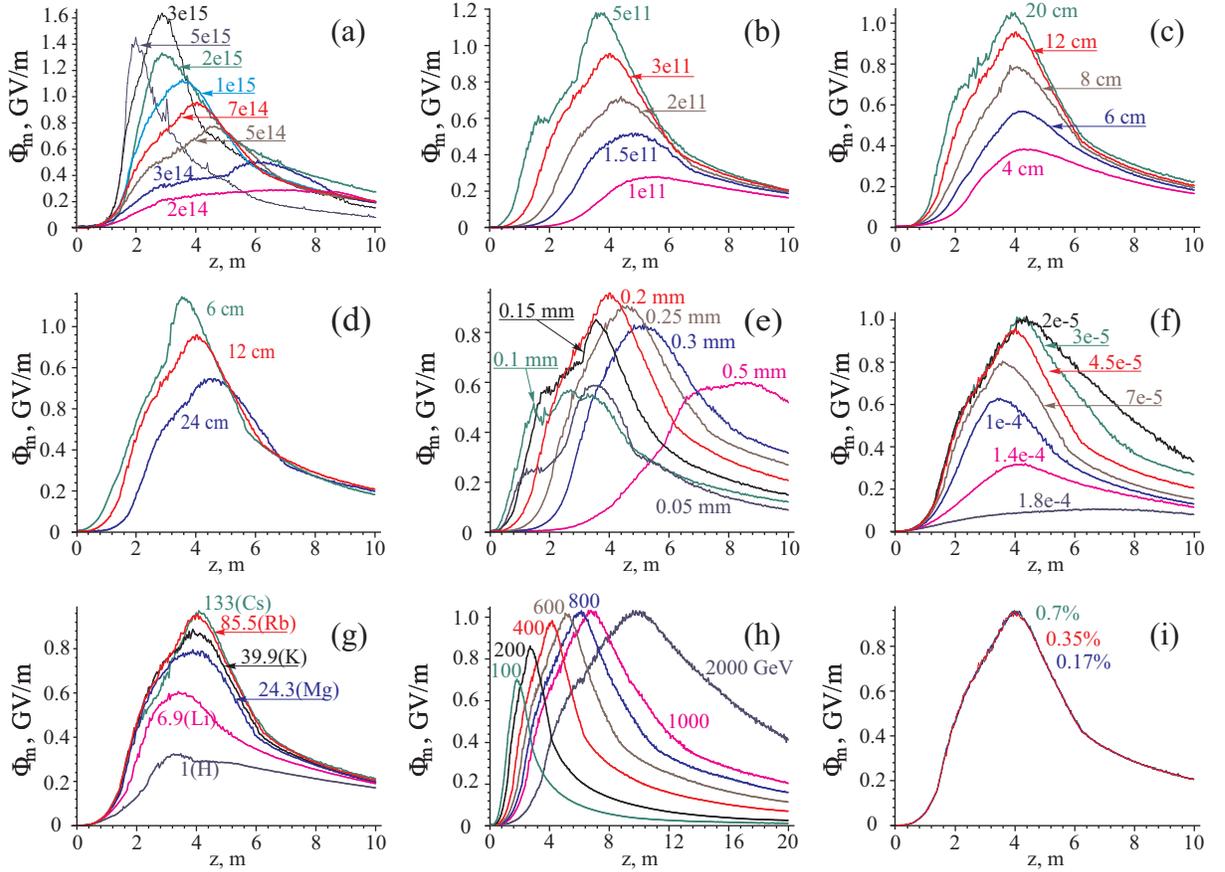} \ec
 \vspace*{-5mm}
\caption{ Length dependencies of the maximum wakefield amplitude $\Phi_m (z)$ grouped by the varied parameter: (a) plasma density $n_0$, (b) beam population $N_b$, (c) beam length at constant peak current $\sigma_{zb}$, (d) beam length at constant population $\sigma_{zb}^*$, (e) beam radius $\sigma_{rb}$, (f) beam angular spread $\delta \alpha_b$, (g) atomic weight of plasma ions $M_i$,   (h) beam energy $W_b$, (i) beam energy spread $\delta W_b$. Values of the variable parameter are indicated near the graphs. }\label{f1-main}
\end{figure*}

\begin{table*}[t]
 \caption{ Some parameters of TeV-class proton accelerators (from Ref.\,\cite{PRD86-010001}).}\label{t2}
 \bc\begin{tabular}{lcccccc}\hline
   & SPS (AWAKE) \ & \ LHC-2009 \ & \ LHC-nominal \ & \ HERA \ & \ TEVATRON \ & \ RHIC \\
  Beam energy (TeV) & 0.4 & 3.5 & 7 & 0.92 & 0.98 & 0.25 \\
  Bunch length (cm) & 12 & 9 & 7.5 & 8.5 & 50 & 70 \\
  Bunch population (units of $10^{11}$) & 3 & 1.5 & 1.15 & 0.7 & 0.9 & 1.65 \\
  Peak current (A) & 50 & 30 & 30 & 15 & 3.5 & 4.5 \\
%  Normalized emittance (mm mrad) & 3.6 & 2.45 & 3.5 & 5 & 3 & 3.75 \\
  \hline
 \end{tabular}\ec
\end{table*}

The key beam parameter determining the wakefield behavior is the peak current or, equivalently, the field increment due to a single micro-bunch focused to $\sigma_{rb} \sim c/\omega_p$ \cite{PoP18-103101}. This parameter does not vary much in modern TeV-class proton accelerators (Table~\ref{t2}). Thus the qualitative results obtained for the SPS proton beam are also applicable to several other machines.

The simulations are made with the quasi-static axisymmetric 2d3v code LCODE \cite{PoP5-785,PRST-AB6-061301,IPAC13-1238,LCODE}. We use cylindrical coordinates $(r, \varphi, z)$ with the $z$-coordinate measured from the plasma entrance and the co-moving coordinate $\xi = z-ct$ measured from the beam central plane.

\section{Wakefield amplitude} \label{s2}

When we analyze the wakefield amplitude, we refer to the maximum $\Phi_\text{max} (z)$ of the dimensionless wakefield potential $\Phi (z,\xi)$ on the axis:
\begin{equation}\label{e1}
    \Phi (z,\xi) = \frac{\omega_p}{c E_0} \int_\xi^{\infty} E_z(z, \xi') \, d \xi',
\end{equation}
where $E_z$ is the on-axis electric field. The maximum is taken over all values of $\xi$ for a fixed $z$. The reason for using this quantity is that the wakefield potential is more noise-resistant than $E_z$ itself. We also use the dimensional quantity $\Phi_m = E_0 \Phi_\text{max}$ to characterize the accelerating gradient.

Main results of the parameter scan are shown in Fig.\,\ref{f1-main}. In each group of graphs we vary one parameter from Table~\ref{t1} and plot functions $\Phi_m (z)$.

\begin{figure}[thb]
\bc\includegraphics[width=223bp]{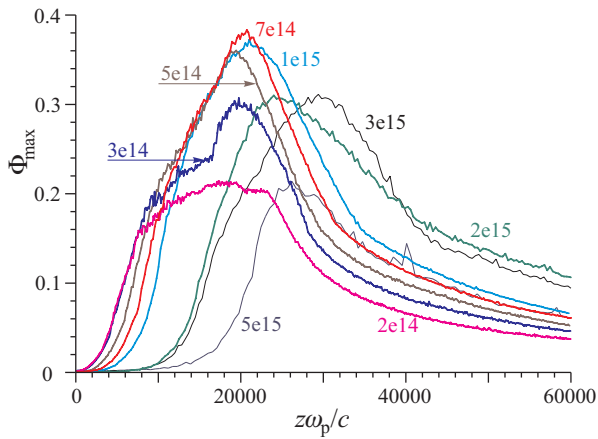} \ec
 \vspace*{-5mm}
\caption{The dimensionless wakefield amplitude $\Phi_\text{max}$ versus the dimensionless propagation distance $z \omega_p/c$ for plasmas of various densities. Coloring of the graphs is the same as in Fig.\,\ref{f1-main}(a). }\label{f2-density}
\end{figure}
The plasma density dependence of the wakefield amplitude [Fig.\,\ref{f1-main}(a)] is mainly determined by the scaling $E_0 \propto \sqrt{n_0}$. Namely, the higher the plasma density is, the stronger field this plasma can support. At high plasma densities, the beam is several plasma skin depths wide, and beam filamentation could develop \cite{PF30-252}. This phenomenon is substantially three-dimensional and cannot be simulated by the axisymmetric code. The exact beam width at which filamentation starts to dominate over self-modulation is not clear yet. We therefore plot the highest density curves by thinner lines to indicate that these curves may not be realistic.

The level of wakefield saturation can be understood from dependence of the dimensionless amplitude $\Phi_\text{max}$ on the dimensionless propagation distance (Fig.\,\ref{f2-density}). We see that there is an optimum plasma density at which the dimensionless wakefield amplitude is maximal. At higher densities, the wakefield is limited by nonlinear elongation of the plasma wave period. The approximate theory of this effect \cite{PoP20-083119} gives the following expression for the maximum wakefield amplitude at the moment of optimum beam micro-bunching:
\begin{equation}\label{e2}
    \Phi_\text{max} = \left( \frac{4 \beta \Delta E}{\alpha E_0} \right)^{1/3},
\end{equation}
where $\Delta E$ is the wakefield increment due to one micro-bunch, and $\alpha \approx 0.1$ and $\beta \approx 0.5$ are numerical factors that depend on the beam radius. Formula (\ref{e2}) gives $\Phi_\text{max} \approx 0.4$ for the baseline case in agreement with Fig.\,\ref{f2-density}. Since the plasma wave is weakly nonlinear and the beam is wider than $c/\omega_p$, the contribution $\Delta E$ of a single micro-bunch is determined mainly by the beam density \cite{PoP18-103101}. Thus, $\Delta E$ changes almost not at all if we change the plasma density. The limiting field (\ref{e2}) depends on the plasma density through $E_0$ and $\alpha$. Compared to the baseline variant ($n_0 = 7 \times 10^{14} \text{cm}^{-3}$), the higher density variant with $n_0 = 5 \times 10^{15} \text{cm}^{-3}$ has approximately 2.7 times higher $E_0$, twice bigger $\alpha$ (as the wave is almost one dimensional with $\alpha=3/16$), and 40\% smaller expression (\ref{e2}) for $\Phi_\text{max}$ in good agreement with Fig.\,\ref{f2-density}. The wakefield amplitude at high densities is thus limited by the nonlinear saturation of wave growth.

The theory \cite{PoP20-083119} also specifies the number of micro-bunches coherently exciting the wakefield:
\begin{equation}\label{e3}
    N_\text{eff} = \left( \frac{4 \beta E_0^2}{\alpha \Delta E^2} \right)^{1/3}.
\end{equation}
Multiplied by the plasma wavelength $2\pi c/\omega_p$, this number gives the length $L_\text{eff}$ of the beam part that efficiently excites the wave. For the baseline variant $N_\text{eff} = 130$ and $L_\text{eff} = 16$\,cm, which is roughly the whole available beam. At smaller plasma densities the plasma wavelength is longer, the number $N$ of macro-bunches in the beam is smaller than (\ref{e3}), the field growth along the beam does not come to saturation, and the maximum field is roughly $N \Delta E \propto \sqrt{n_0}$. The low-density curves in Fig.\,\ref{f2-density} follow this scaling quite well, as the curve maxima for $n_0 = (5,3,2) \times 10^{14} \text{cm}^{-3}$ constitute a ratio of $\sqrt{5} : \sqrt{3} : \sqrt{2}$. The wakefield amplitude at low plasma densities is thus limited by the beam length, and the baseline case is located close to the border between plasma nonlinearity-limited and beam length-limited regimes of field excitation.

\begin{figure}[b]
\bc\includegraphics[width=219bp]{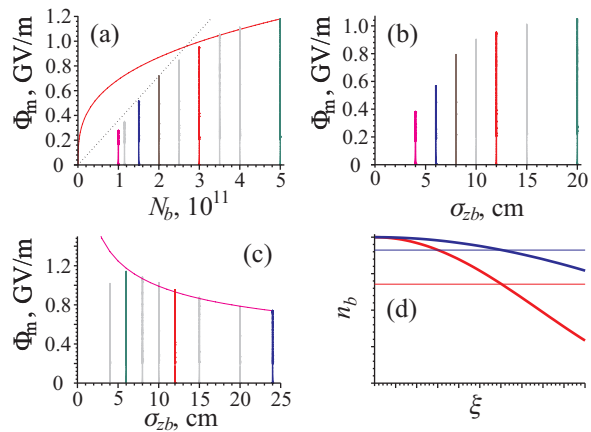} \ec
 \vspace*{-5mm}
\caption{The absolute maximum of the wakefield amplitude $\Phi_m$ (bar height) versus varied parameters: (a) beam population for the fixed beam length, (b) beam length for the fixed beam peak current, and (c) beam length for the fixed beam population. Coloring of the bars is the same as in Fig.\,\ref{f1-main}(b-d). Illustration of true (thick lines) and effective (thin lines) beam densities for normal (red) and elongated (blue) beams (d). }\label{f3-population}
\end{figure}
The dependence of the wakefield amplitude on the beam population, either for the fixed beam length [Fig.\,\ref{f1-main}(b)] or for the fixed peak current [Fig.\,\ref{f1-main}(c)], is also determined by interplay of the above two limitations. To demonstrate this, we plot the maximum wakefield amplitude versus varied parameters (Fig.\,\ref{f3-population}). For the fixed beam length, $\Delta E \propto N_b$, and the formula (\ref{e2}) gives the scaling $\Phi_m \propto \Phi_\text{max} \propto N_b^{1/3}$ [thin solid line in Fig.\,\ref{f3-population}(a)]. If the wakefield amplitude is limited by the beam length, then it is directly proportional to the number of particles driving the wave, and $\Phi_m \propto N_b$ (thin dotted line). The baseline case (always plotted in red) is in the transition region.

For the fixed beam current there is no simple scaling for the maximum amplitude in the nonlinearity-limited regime. For a uniform beam with a constant current we may expect no dependence on the beam length. Here the beam has a Gaussian-like density distribution, so the effective beam density is smaller than the peak density. The longer the beam, the closer the effective density to the peak density [Fig.\,\ref{f3-population}(d)]. As the average $\Delta E$ is proportional to the effective density, we observe a weak amplitude growth with the beam population in the nonlinearity-limited part of Fig.\,\ref{f3-population}(b). In the length-limited part, $\Phi_m \propto N_b \propto \sigma_{zb}$.

If we change the beam length keeping the beam population fixed [Fig.\,\ref{f1-main}(d)], then again the result follows the scaling~(\ref{e2}). Now $\Delta E \propto n_{b0} \propto \sigma_{zb}^{-1}$, and $\Phi_m \propto \sigma_{zb}^{-1/3}$. This scaling is shown in Fig.\,\ref{f3-population}(c) with a thin line. For small $\sigma_{zb}$ the beam length decreases faster than $N_\text{eff} \propto \sigma_{zb}^{2/3}$, thus short beams are in the length-limited regime with $\Phi_m\approx \text{const}$.

\begin{figure}[b]
\bc\includegraphics[width=214bp]{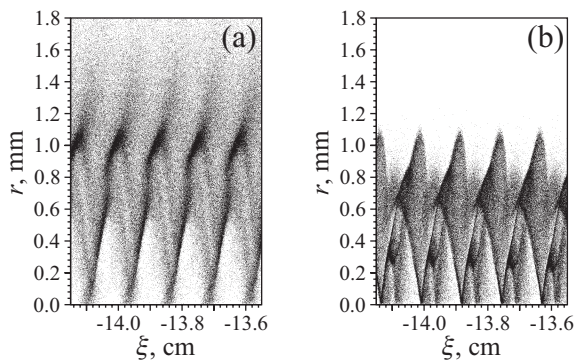} \ec
 \vspace*{-5mm}
 \caption{ Real space portraits of beam fragments at the time of developed modulation ($z=3.6$\,m) for initial beam radius 0.05\,mm (a) and 0.2\,mm (b).}\label{f4-divergence}
\end{figure}
As we can see from Fig.\,\ref{f1-main}(e), the baseline beam radius is the optimum one for the specified set of other parameters. This optimum appears due to two effects. At large beam radii the contribution $\Delta E$ of a micro-bunch drops as $\sigma_{zb}^{-1}$. The total wakefield drops accordingly, since wave excitation for $\sigma_{zb}>0.2$\,mm is in the length-limited regime. At small radii the beam has a large angular spread that causes fast beam divergence and reduction of the maximum wakefield amplitude \cite{PoP18-103101}. The diffuse character of beam divergence for small $\sigma_{zb}$ is illustrated with Fig.\,\ref{f4-divergence} in comparison with the usual self-modulation.

The dependence of the wakefield amplitude on the initial angular spread [Fig.\,\ref{f1-main}(f)] is determined by emittance driven divergence. If the emittance is below some threshold value, it has no effect on the amplitude and affects only the rate of beam degradation at the full modulation stage \cite{PoP18-024501}. As the emittance exceeds the threshold, the wakefield amplitude steeply drops down. Thus the baseline variant has no safety margin for beam emittance, and a slightly higher emittance would cause substantial degradation of the wakefield.

The dependence of the maximum wakefield on the ion atomic weight [Fig.\,\ref{f1-main}(g)] is in line with results of Ref.\cite{PRL109-145005}. For the chosen beam parameters, reduction of the wakefield due to ion motion is substantial for atomic weights below 40, but there is no large safety margin. For improved beam parameters, ion motion may be an issue.

\begin{figure}[tb]
\bc\includegraphics[width=191bp]{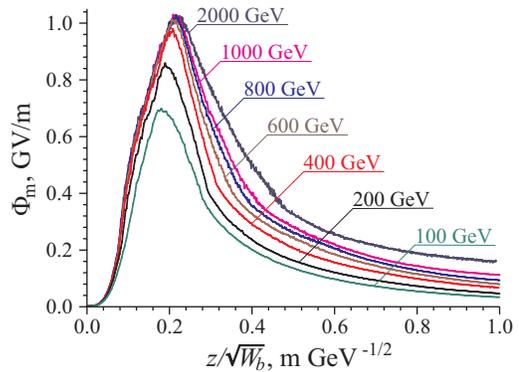} \ec
 \vspace*{-5mm}
\caption{ The maximum wakefield amplitude $\Phi_m$ versus the normalized propagation distance $z/\sqrt{W_b}$ for various beam energies. }\label{f5-energy}
\end{figure}
The beam energy has a small effect on the wakefield amplitude [Fig.\,\ref{f1-main}(h)], but changes the time scale of the process. The theory \cite{PRL104-255003} predicts that the time scale of self-modulation is proportional to $\sqrt{W_b}$, and this scaling is nicely reproduced in simulations (Fig.\,\ref{f5-energy}). Visible deviations from the common curve shape in Fig.\,\ref{f5-energy} observed for lowest and highest energies are due to the emittance change that accompanies the energy variation. At low energies the amplitude is reduced by emittance-driven divergence, at high energies the amplitude decreases slower at the stage of beam degradation.

Reasonable beam energy spreads have no effect on the wakefield [Fig.\,\ref{f1-main}(i)].

\begin{figure}[htb]
\bc\includegraphics[width=174bp]{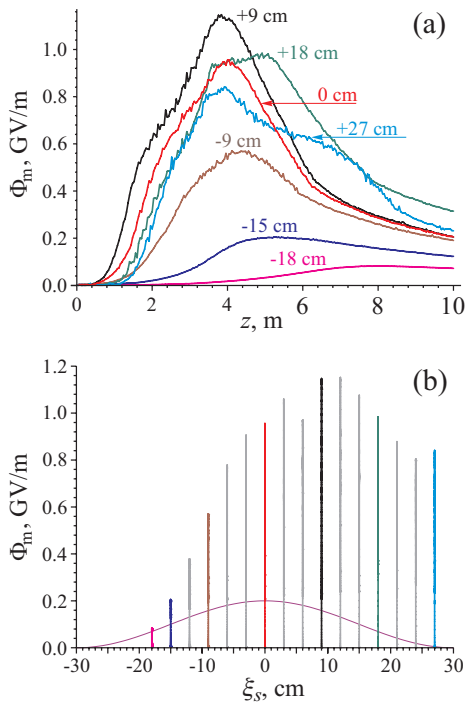} \ec
 \vspace*{-5mm}
\caption{ (a) Length dependencies of the maximum wakefield amplitude $\Phi_m (z)$ for various positions $\xi_s$ of the beam leading edge (marked near the curves) and (b) the absolute maximum of the wakefield amplitude $\Phi_m$ (bar height) versus the position of the leading edge. The curve in fragment (b) shows the shape of uncut beam. }\label{f6-timing}
\end{figure}
The wakefield amplitude also depends on the position $\xi_s$ of the ionizing laser pulse with respect to the beam central plane (Fig.\,\ref{f6-timing}). This distance is controlled by time synchronization of laser and proton beams. As propagation velocities of both beams are close to the speed of light, this distance changes almost not at all during the development of self-modulation. Therefore the plasma response to the two beams (proton and laser ones) is the same as to the single proton beam with a hard leading edge located at $\xi_s$. We simulate the latter case and take the initial proton beam density in the form
\begin{multline}\nonumber
 n_b (r, \xi) = 0.5 \, n_{b0} \, e^{-r^2/2 \sigma_{rb}^2} \left[  1 + \cos \left(  \sqrt{\frac{\pi}{2}} \frac{\xi}{\sigma_{zb}}  \right)  \right], \\
  -\sigma_{zb} \sqrt{2\pi} < \xi < \xi_s,
\end{multline}
which is close to Gaussian distribution and smoothly vanishes at $|\xi| = \sigma_{zb} \sqrt{2\pi}$. For negative $\xi_s$ the wakefield is expectedly small, as there is not enough charge in the intact part of the beam. The strongest wakefield is observed for $\xi_s \approx \sigma_{zb}$, as in this case the seed perturbation is strong enough and the most dense part of the beam undergoes self-modulation. A similar result was earlier obtained for self-modulated electron beams \cite{PoP19-063105}. A smaller wakefield amplitude for $\xi_s > \sigma_{zb}$ is due to smaller seed perturbations and later development of self-modulation [Fig.\,\ref{f6-timing}(a)].

To summarize the obtained results, we list sensitivities $S_a$ of the wakefield amplitude to variation of beam and plasma parameters in Table~\ref{t3}. We define $S_a$  as the proportionality coefficient between the relative variation of the absolute maximum of $\Phi_m$ and the small relative variation of a quantity $X$:
\begin{equation}\label{e4}
    \delta \Phi_m /  \Phi_m = S_a^X \delta X / X.
\end{equation}
To calculate the seed location sensitivity, we take $\sigma_{zb}$ as $X$. As we can see, there are no parameters of the system the wakefield amplitude is extremely sensitive to.

\section{Wakefield phase} \label{s3}

\begin{figure}[bt]
\bc\includegraphics[width=220bp]{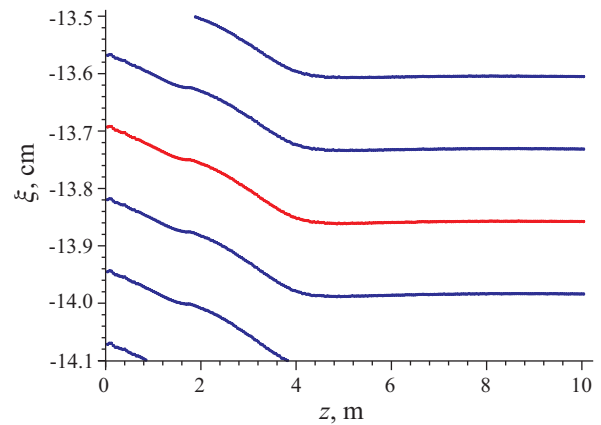} \ec
 \vspace*{-5mm}
\caption{ Co-moving coordinates $\xi_m$ of several local maxima of the wakefield potential versus the propagation distance $z$ for the baseline variant. The maximum used for sensitivity calculation is shown in red.}\label{f7-phase}
\end{figure}

The wakefield phase established after self-modulation is also of high importance as it shows  the optimal location for the accelerated particles. A typical phase behavior is illustrated by Fig.\,\ref{f7-phase}. Here we show how local maxima $\xi_m$ of the wakefield potential $\Phi$ moves in the co-moving frame during beam self-modulation and subsequent degradation of micro-bunches. The slope of these curves determines the local phase velocity $v_\text{ph}$ of the plasma wave:
\begin{equation}\label{e5}
    v_\text{ph} = c \left(1 + \frac{d \xi_m}{dz} \right).
\end{equation}
The interval $-14.1\,\text{cm} < \xi < -13.5\,\text{cm}$ is selected for presentation since this is the place where the bunch of  test electrons must be injected to produce a narrow final energy spectrum \cite{AWAKE}. This figure explains the choice of the optimum injection parameters given in Ref.\,\cite{AWAKE}. The optimum electron energy 16\,MeV is that for which the electron velocity equals $v_\text{ph}$ at the stage of self-modulation ($z<4$\,m). The optimum injection place ($z \approx 4$\,m) corresponds to the length of full micro-bunching. The injection delay $\xi \approx 13.8$\,cm with respect to the laser pulse corresponds to the area where the phase line flattens out immediately after self-modulation has developed.

\begin{figure}[tbh]
\bc\includegraphics[width=230bp]{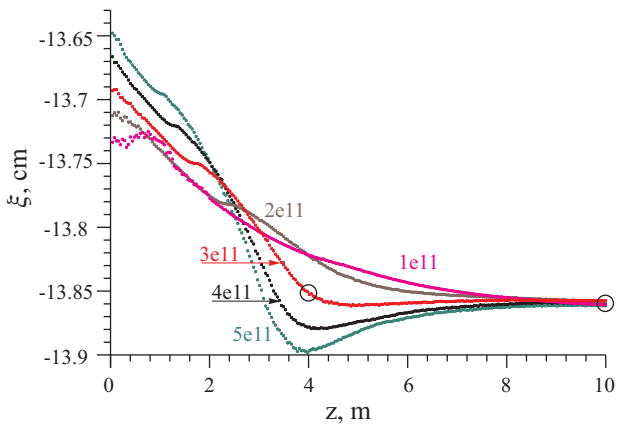} \ec
\caption{ A family of constant phase curves for various beam populations. Circles denote the points used for calculation of phase sensitivities. }\label{f8-phases}
\end{figure}

\begin{table}[tb]
 \caption{ Sensitivity of wakefield amplitude and phase to variations of beam and plasma parameters.}\label{t3}
 \bc\begin{tabular}{lccc}\hline
  Parameter &  $S_a$  &  $S_\varphi$ (4\,m)  &  $S_\varphi$ (10\,m) \\ \hline
  Plasma density & 0.4 & 375 & 340 \\
  Beam population & 0.7 & -4.75 & 0.026 \\
  Beam length (fixed peak current) & 0.3 & -0.5 & -0.19 \\
  Beam length (fixed population) & -0.4 & 4.2 & -0.25 \\
  Beam radius & 0 & 2.5 & -0.42 \\
  Beam angular spread & -0.3 & 0.9 & 0.26 \\
  Atomic weight of plasma ions & 0 & 0.85 & 0.1 \\
  Beam energy & 0.07 & 1.9 & 0.38 \\
  Beam energy spread & 0 & 0 & 0 \\
  Seed location & 0.3 & 1.3 & -0.39 \\
  \hline
 \end{tabular}\ec
\end{table}

To quantify the sensitivity of wakefield phase to parameter variations, we introduce the phase sensitivity $S_\varphi$ as
\begin{equation}\label{e6}
    \delta \xi_m \omega_p / c = S_\varphi^X \delta X / X,
\end{equation}
where $\delta \xi_m$ is the shift of some constant phase point. The phase curves may change in a complicated way due to variations of beam or plasma parameters, as exemplified by Fig.\,\ref{f8-phases}. For most curves, the strongest phase deviation is located near the amplitude maximum (at $z \sim 4$\,m), while at the end of the plasma section the phase shift is small. We therefore calculate $S_\varphi$ in two plasma cross-sections: $z=4$\,m and $z=10$\,m. To this end we take the local maximum of $\Phi (\xi)$ located at $\xi\approx -13.86$\,cm.

Phase sensitivities listed in Table~\ref{t3} display several noticeable features. The most flaring one is the high sensitivity to plasma density variations. As shown in Ref.\,\cite{PoP20-013102}, a small density variation $\delta n_0$ results in the relative change $\delta \lambda_p / \lambda_p = - \delta n_0 / (2 n_0)$ of the plasma wavelength and the forward shift of the wakefield pattern by $L_\text{per} \delta n_0 / (2 n_0)$, where $L_\text{per}$ is the distance between the observation point and the area of wakefield formation. This distance was not defined strictly in Ref.\,\cite{PoP20-013102}, and the phase sensitivity can be used to refine the definition. Here $L_\text{per} = 2 S_\varphi^{n_0} c/\omega_p \approx 14$\,cm, so the full distance to the seed laser pulse is the ``effective'' wakefield length if the beam self-modulates in the plasma of a detuned density. This contrasts to the case considered in Ref.\,\cite{PoP20-013102}, where an already modulated beam enters a perturbed density plasma, and $L_\text{per}$ is roughly twice shorter.

The extreme phase sensitivity to plasma density variations makes it challenging to deterministically inject externally generated short electron bunches into the wakefield of self-modulating proton beam. Assume that the bunch must be placed into a certain accelerating bucket with a longitudinal precision of about $0.1\,c/\omega_p$. Then it straightforwardly follows that the plasma density must be held to a designated value within $0.1/S_\varphi^{n_0} \approx 0.03$\%.

Another noticeable feature is that all other sensitivities are small at $z=10$\,m. Values less than unity means that even a twofold change of a parameter moves a fixed phase point by a small fraction of the wakefield period. Thus the wakefield structure established after partial destruction of micro-bunches is well phase locked to the seed perturbation.

At $z=4$\,m sensitivities to beam population, beam length at fixed current, and beam radius are high since these parameters determine the beam density $n_{b0}$ which in turn governs the theoretically predicted phase velocity $v_\text{ph}^\text{th}$ of the growing self-modulation mode \cite{PRL107-145003}:
\begin{equation}\label{e7}
    1-\frac{v_\text{ph}^\text{th}}{c} \approx \frac{1}{2} \left( \frac{\xi}{z} \right)^{1/3} \left( \frac{n_{b0} m c^2}{2 n_0 W_b} \right)^{1/3}.
\end{equation}
From (\ref{e7}) we may expect $(S_\varphi^{N_b}, S_\varphi^{\sigma_{zb}}, S_\varphi^{\sigma_{rb}},  S_\varphi^{W_b})$ to relate as $(-2, 2, 1, 2)$. This reasonably agrees with the simulations except for the sensitivity to the beam energy which is roughly twice lower.

Note also the relatively high sensitivity to the atomic weight of plasma ions at $z=4$\,m. In comparison to the amplitude sensitivity this shows that ion motion manifests itself much stronger in phase shifts than in the amplitude change.

\section{Discussion} \label{s4}

The performed study has not only clarified the sensitivity of the excited wakefield to beam and plasma parameters (summarized in Table~\ref{t3}), but also helped to identify physical effects limiting the wakefield amplitude in different parts of the parameter space. The most important effect is the nonlinear elongation of the wakefield period \cite{PoP20-083119}. As long as this effect has not come into play, the wakefield amplitude is directly proportional to the number of protons driving the wakefield. Once the limit is achieved by improving some of the parameters, further growth of the accelerating field drastically slows down. Two other important effects are emittance driven divergence and motion of plasma ions. Both effects are characterized by some threshold values or, to be exact, surfaces in the multi-dimensional parameter space. If the threshold is crossed, the wakefield amplitude rapidly drops. If not, there is no influence on the wakefield amplitude.

The baseline parameter set of the AWAKE experiment (Table~\ref{t1}) falls exactly at the onset of  nonlinear period elongation and close to thresholds of emittance driven divergence and ion motion. In other words, this is the point beyond which efforts in increasing beam charge or peak current will not result in the proportional increase of the wakefield amplitude. Also there is no much safety margin in beam emittance and ion weight.
A way to stronger wakefields may probably be opened by operating at lower beam emittances and higher plasma densities, but this is a subject of a separate study.

Very high sensitivity of the wakefield phase to plasma density variations stimulates interest in injection methods based on selective trapping of electrons from an initially long (several wavelengths) beam \cite{PRL107-145003,JPP78-455}.

\section{Acknowledgements}

The authors are grateful to participants of AWAKE collaboration for stimulating discussions.

This work was supported by The Ministry of Education and Science of the Russian Federation, Siberian Supercomputer Center SB RAS, and RFBR grant 14-02-00294.

\end{document}